\documentclass{PoS}

\title{Interpretation of the unpolarized azimuthal asymmetries in SIDIS}

\ShortTitle{Interpretation of the unpolarized azimuthal asymmetries in SIDIS}

\author{Albi Kerbizi$\,^{*}$ \\
        \rm{\textbf{On Behalf of the COMPASS Collaboration}}\\
        $^{*}\,$University of Trieste and INFN\\
        E-mail: \email{albi.kerbizi@ts.infn.it}}

\abstract{The measurement of azimuthal modulations in hadron leptoproduction on unpolarized nucleons allows to get information on the intrinsic transverse momentum $\langle k_T^2 \rangle$ of quarks in a nucleon through both the Cahn effect and the Boer-Mulders function.

We have compared the azimuthal asymmetries in the cross section of $160\, \rm{GeV}/c$ muons scattered off an unpolarised deuteron target as measured by COMPASS with a Monte Carlo program, based on the ${}^3P_0$ model of quark anti-quark pair production at string breaking, which accounts for the Cahn effect.
Large differences have been observed between data and Monte Carlo, in particular at large values of the fraction of the longitudinal momentum of the fragmenting quark carried by the produced hadron. We found out that most of these differences are due to pions from exclusive vector mesons contaminating the SIDIS sample, which also exhibit large azimuthal modulations. Using the measurements of the exclusive reaction $\mu N\rightarrow \mu' \, \rho\, N$ we had done in 2006, we can reproduce reasonably well the observed differences.

Subtracting the contribution of hadrons produced in the decay of exclusive vector mesons from the SIDIS unpolarised azimuthal asymmetries is therefore a prerequisite condition for extracting $\langle k_T^2 \rangle$ and the Boer-Mulders function.}

\FullConference{23rd International Spin Physics Symposium - SPIN2018 -\\
		10-14 September, 2018\\
		Ferrara, Italy}

\begin{document}

\section{Introduction}

The azimuthal asymmetries in unpolarized Semi-Inclusive Deep Inelastic Scattering (SIDIS) are a powerful tool to access the quark intrinsic transverse momentum $k_T$ and the Boer-Mulders Transverse Momentum Dependent Parton Distribution Function (TMD PDF) $h_1^{\perp q}$. In particular the quantity $A^{UU}_{\cos\phi_h}$, defined as the amplitude of the $\cos\phi_h$ modulation ($\phi_h$ is the azimuthal angle of the hadron with respect to the lepton scattering plane) gives a direct access to $\langle k_T^2 \rangle$ through the Cahn effect, which is the main contributor to $A^{UU}_{\cos\phi_h}$. The situation is therefore more favourable than in the case of the hadron multiplicities as function of the hadrons transverse momentum $p_T^h$, which probe a combination between both the quark intrinsic $k_T$ and the transverse momentum $p_{\perp}$ the hadrons acquired in the fragmentation process. The asymmetry $A^{UU}_{\cos 2\phi_h}$, which is the amplitude of the $\cos 2\phi_h$ modulation, gives access to the Boer-Mulders TMD PDF.

Recent measurements of the unpolarized SIDIS azimuthal asymmetries have been provided by the COMPASS Collaboration \cite{Adolph:2014pwc} on deuteron for charged hadrons, and on both deuteron and proton for charged hadrons, pions and kaons by the HERMES Collaboration \cite{Airapetian:2012yg}.
They all show strong dependencies on the kinematic variables.
Several phenomenological analyses (for more details see Ref. \cite{Barone:2015ksa}) did not succeed either in reproducing the data or in extracting the Boer-Mulders PDF. As a result the present knowledge of the quark intrinsic transverse momentum has very large uncertainties and a possible non-zero Boer-Mulders function in the SIDIS cross-section has still to be demonstrated.

\begin{figure}[h]
	\centering
	\includegraphics[width=0.9\textwidth]{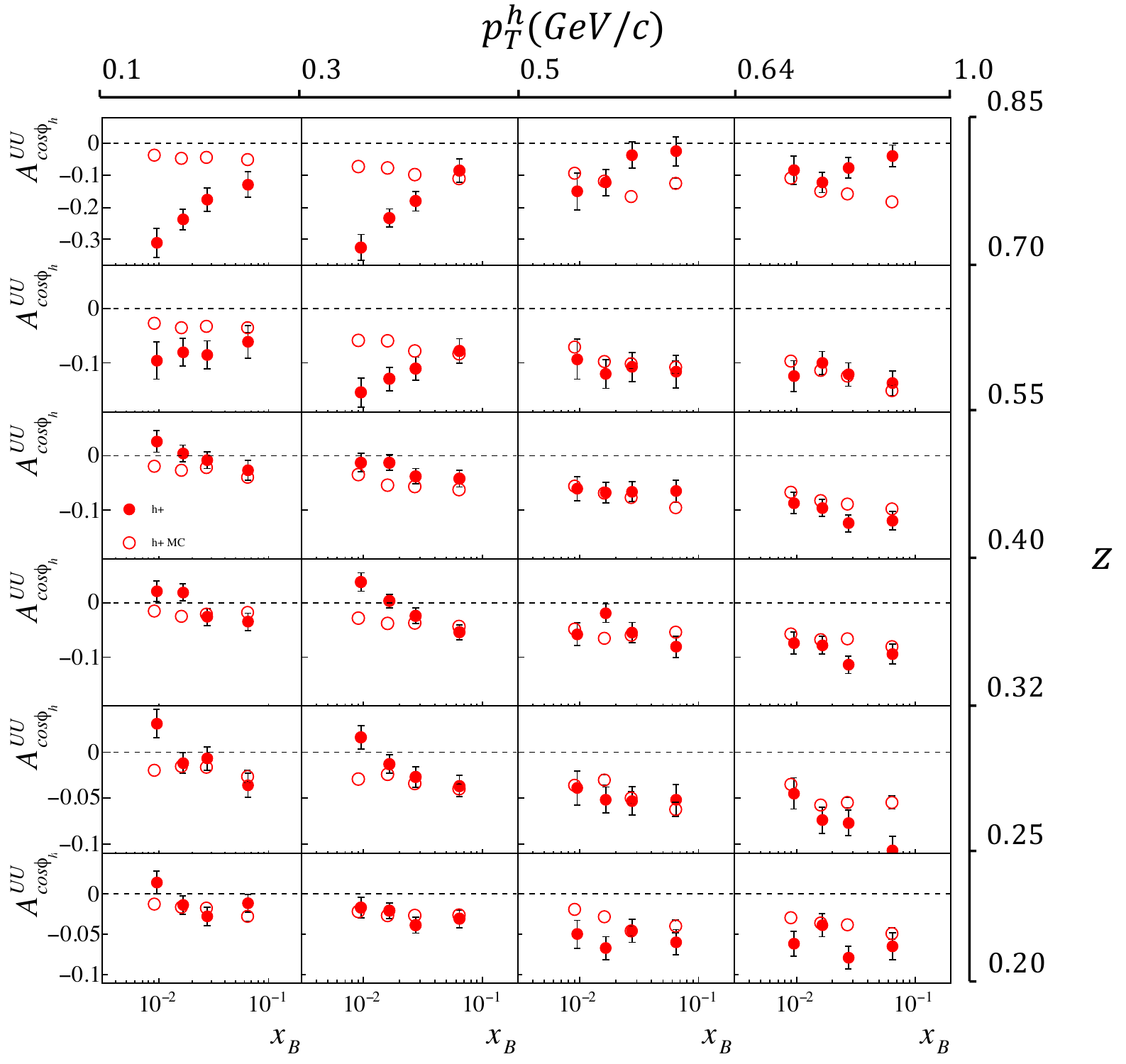}
\caption{Comparison between MC (open points) and published asymmetries from the COMPASS Collaboration \cite{Adolph:2014pwc} for positive hadrons (full points).}
\label{fig:cosphi3dMCpos}
\end{figure}

For the interpretation of the $A^{UU}_{\cos \phi_h}$ asymmetry we used a recently developed MC describing the fragmentation of polarized quarks~\cite{Kerbizi:2018qpp,AlbiDubna}, modified to include the Cahn effect.
The Cahn effect is taken into account by modulating the fragmenting quark direction according to the lepton-quark hard cross section calculated for a non-zero $k_T$.
The $\langle p_{\perp}^2\rangle$ dependence on the fraction $z$ of the quark energy carried by the hadron, is built in as a consequence of the underlying string fragmentation framework of the model and a suitable dependence of $k_{T}^2$ on $x_B$, the Bjorken variable, has been used to reproduce the values of $A^{UU}_{\cos \phi_h}$ at $z\lesssim 0.5$.
The comparison between the MC (open points) and the published COMPASS $A^{UU}_{\cos \phi_h}$ asymmetry \cite{Adolph:2014pwc} (full points) is shown in Fig. \ref{fig:cosphi3dMCpos} for $h^+$ as function of $x_B$ in different $z$ and $p_T^h$ bins.
The description in terms of the Cahn effect is rather good at low $z$ but it is clear that the MC and data have a different trend for $z>0.55$. The same is also true for $h^-$. Such discrepancy suggests that a different mechanism is at work in the large $z$ region. From the measurements of hadron multiplicities~\cite{Adolph:2016bga,Adolph:2016bwc,Aghasyan:2017ctw} it is known
that the charged SIDIS sample at large $z$ and at small $p_T^h$ contains many hadrons produced in the decay of exclusive vector mesons (VM) and in particular of $\rho^0$'s.
We have thus investigated the possibility that this VM contribution is at the origin of the discrepancy. For the first time we have measured azimuthal asymmetries for $h^+$ and $h^-$ produced in exclusive events (called "exclusive hadrons" in the following), found them to be large, and subtracted this contribution from the published COMPASS asymmetries. This considerably improves the agreement with the MC.

The article is organized as follows:
the strategy for the determination of the exclusive VM contribution to the unpolarized azimuthal asymmetries is given in Sec. \ref{sec:strategy}. The measurements of the azimuthal asymmetries in the distributions of exclusive hadrons are presented in Sec. \ref{sec:measurement}. In Sec. \ref{sec:percentage} the percentage of hadrons from exclusive events is calculated, whereas Sec. \ref{sec:pureSIDIS} is dedicated to the subtraction of the exclusive hadron contribution in the COMPASS published data. Finally the conclusions are drawn in Sec. \ref{sec:conclusions}.

\section{Strategy for the estimation of the exclusive hadron contribution}\label{sec:strategy}

The SIDIS hadron sample contains unidentified hadrons produced in the DIS regime (defined by $Q^2>1\, (\rm{GeV}^2/c^{2})$ and $W>5\,(\rm{GeV}/c^{2})$ in the COMPASS case). In the standard notation, $Q^2$ is the exchanged photon virtuality and $W$ the final state hadronic mass. From the SIDIS cross section in the one-photon-exchange approximation one expects the azimuthal distribution of the SIDIS hadrons to be
\begin{equation}
    f_h(\phi_h)\propto 1+ \varepsilon_1(y) A^{UU}_{\cos\phi_h} \cos\phi_h+ \varepsilon_2(y) A^{UU}_{\cos 2\phi_h} \cos 2\phi_h,
\end{equation}
where $\epsilon_1=[2(2-y)\sqrt{1-y}]/[1+(1-y)^2]$ and $\epsilon_2=[2(1-y)]/[1+(1-y)^2]$ are kinematic factors (y is the energy fraction of the incoming lepton carried by the exchanged virtual photon).
However, the SIDIS sample contains also events coming from exclusive processes which can not be interpreted in the parton model.

The contribution of exclusive hadrons to $A^{UU}_{\cos\phi_h}$ and $A^{UU}_{\cos 2\phi_h}$ has been estimated following a number of different steps, namely:

\begin{itemize}
\item[a.] The azimuthal distribution of exclusive hadrons is assumed to have the angular dependence
\begin{equation}\label{eq:fexcl}
    f_{\rm{excl}}(\phi_h)=a_0[1+\epsilon_1(y) a_1\cos\phi_h +\epsilon_2(y) a_2\cos 2\phi_h]
\end{equation}
in each $x_B$, $z$ and $p_T^h$ bin of the 3D analysis. Other possible orthogonal modulations are not relevant for this study and consequently have not been considered.

\item[b.] The acceptance-corrected azimuthal distributions of exclusive hadrons are fitted with the function in Eq. (\ref{eq:fexcl}) to obtain the estimates $\hat{a}_1=a^{UU,excl}_{\cos\phi_h}$ and $\hat{a}_2=a^{UU,excl}_{\cos 2\phi_h}$. In the following we indicate the quantities $a^{UU,excl}_{\cos\phi_h}$ and $a^{UU,excl}_{\cos 2\phi_h}$ as the "exclusive amplitudes".

\item[c.] The fraction of exclusive hadrons present in the total data sample, given by
\begin{equation}\label{eq:r}
    r=N_h^{excl}/N_h^{tot},
\end{equation}
where $N_h^{tot}$ is the total number of hadrons present in the sample and $N_h^{excl}$ is the number of hadrons generated in the exclusive processes.
The fraction $r$ is estimated using both HEPGEN and LEPTO MC event generators, as already done for the study of the hadron multiplicities \cite{Adolph:2016bga,Adolph:2016bwc,Aghasyan:2017ctw}. 

\item[d.] Assuming that the exclusive VM production is the only contamination source in the total data sample, the "pure" unpolarized SIDIS azimuthal asymmetries are obtained as
\begin{eqnarray}\label{eq:pure_A}
    A_{\cos i\phi_h}^{UU}|_{VM\, sub.}= \frac{A_{\cos i\phi_h}^{UU}|_{Pub.}-r\,a_{\cos i\phi_h}^{UU,excl}}{1-r}, &&  i=1,2.
\end{eqnarray}
\end{itemize}

\section{Measurement of the azimuthal asymmetries for exclusive events}\label{sec:measurement}

The asymmetries in the azimuthal distributions of exclusive hadrons have been extracted using the SIDIS data sample collected in 2006 by COMPASS on a ${}^6\rm{LiD}$ target, in the same multi-dimentional binning as for the published asymmetries \cite{Adolph:2014pwc}. These data were collected with the same target material and at the same beam energy, thus the exclusive processes can safely be assumed to be the same.

The SIDIS sample is selected with the same cuts as in Ref. \cite{Adolph:2014pwc}, namely
\begin{eqnarray}
\nonumber &&Q^2>1(\rm{GeV}/c)^2,\,\, W>5(\rm{GeV}/c^2),\,\, 0.2<y<0.9, \\
\nonumber  &&\theta^{lab}_{\gamma^*}<60\, \rm{mrad},\,\, 0.003<x_B<0.13,\,\, 0.2<z<0.85, \\
  &&0.1\,(\rm{GeV}/c)<p_T^h<1.\,(\rm{GeV}/c).
\end{eqnarray}

\begin{figure}[h]
\begin{minipage}[c]{0.5\linewidth}
\includegraphics[width=\linewidth]{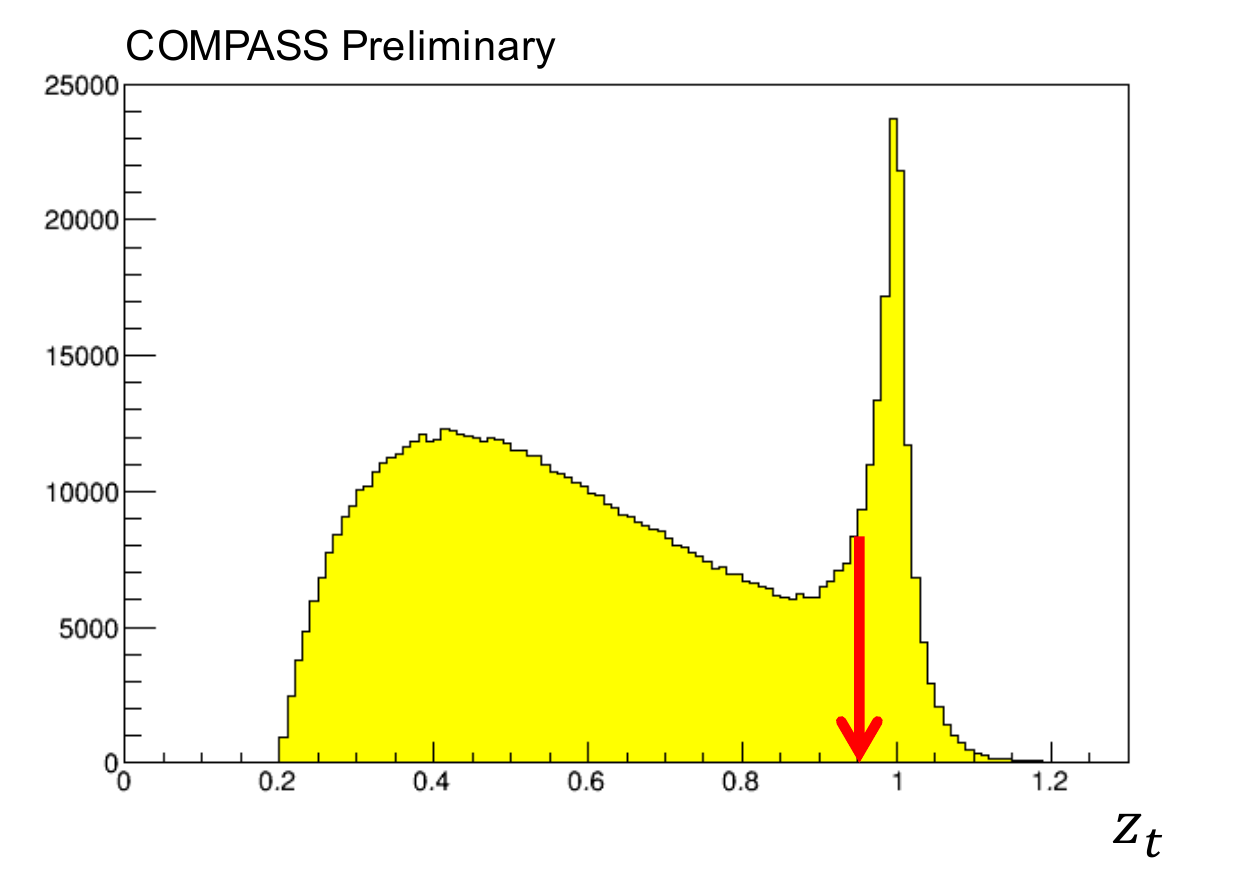}
\end{minipage}
\hfill
\begin{minipage}[c]{0.5\linewidth}
\includegraphics[width=\linewidth]{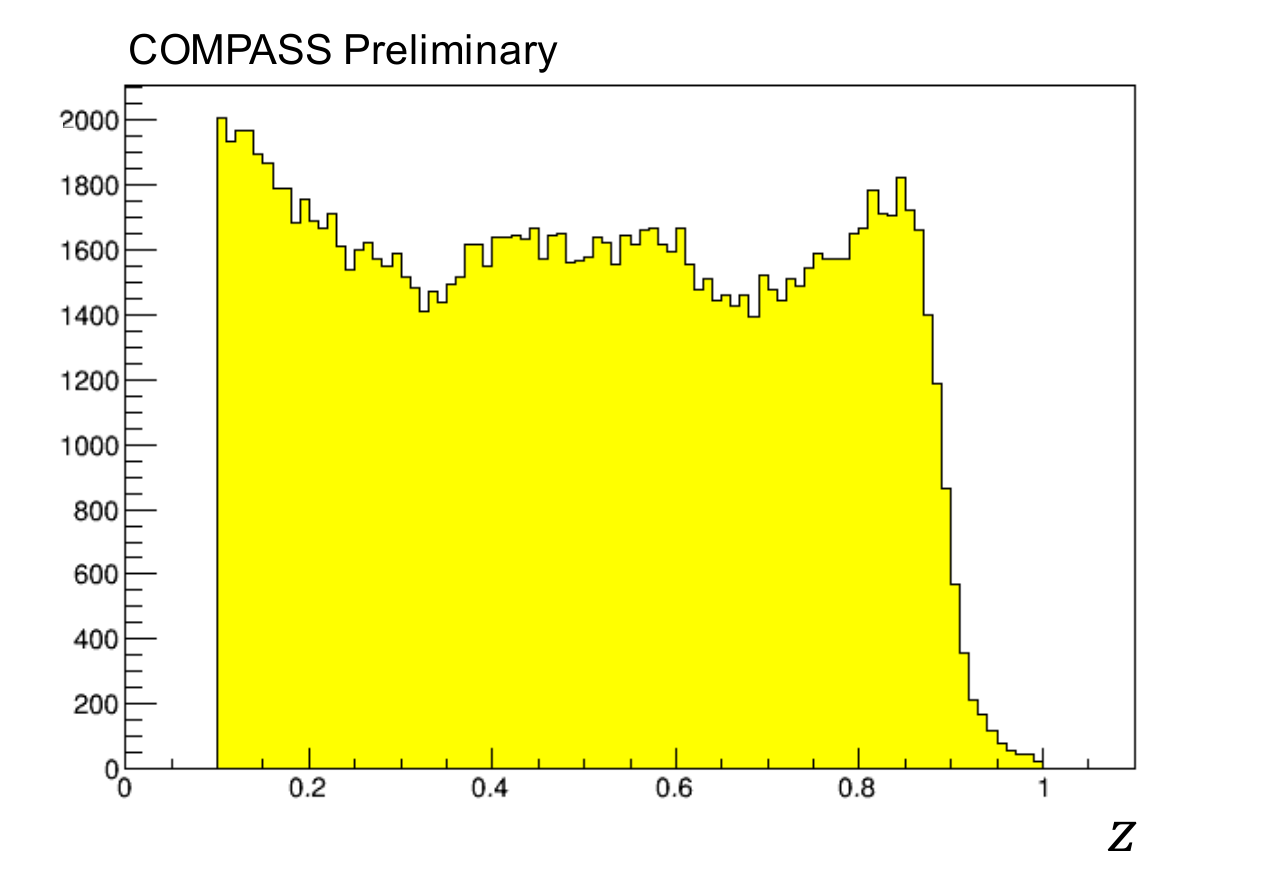}
\end{minipage}%
\caption{In the left plot is shown the distribution of $z_t=z_1 + z_2$ for the events with only a pair of oppositely charged hadrons. The exclusive hadrons are defined by the cut $z_t>0.95$, indicated by the arrow. The $z$ distribution for the positive hadron of the pair is shown in the right plot.}\label{fig:zt}
\end{figure}

\begin{figure}[h]
\begin{minipage}[c]{0.5\linewidth}
\includegraphics[width=\linewidth]{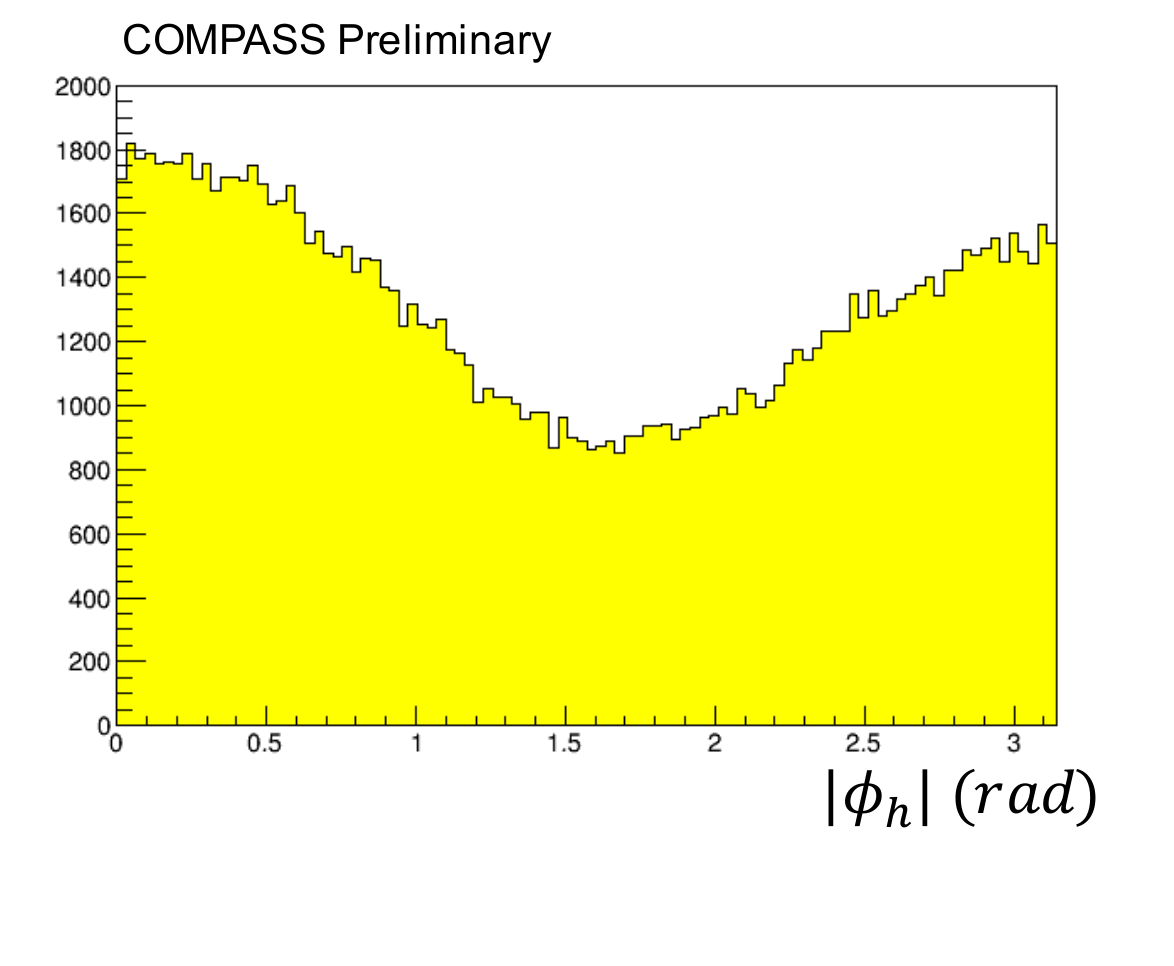}
\end{minipage}
\hfill
\begin{minipage}[c]{0.5\linewidth}
\includegraphics[width=\linewidth]{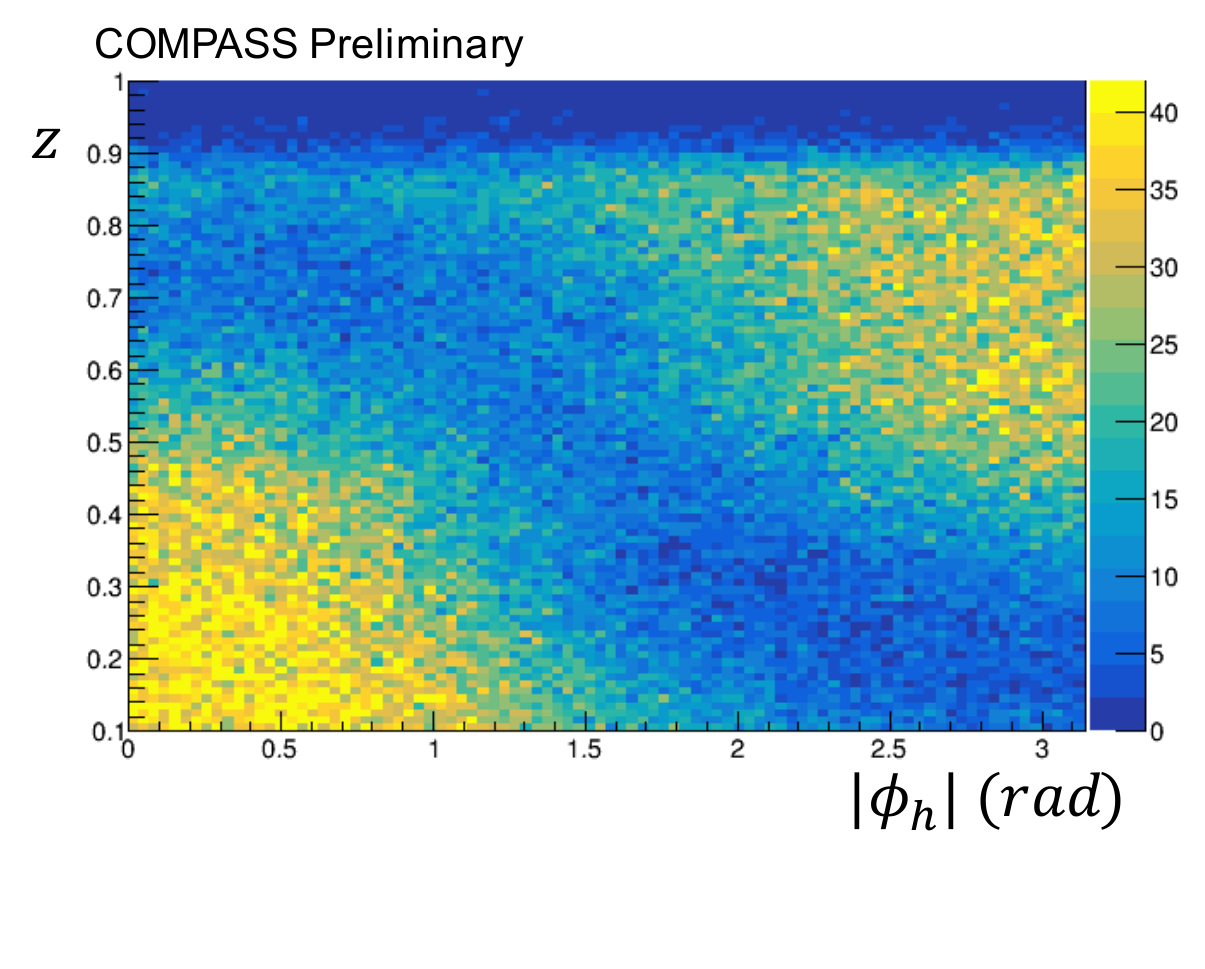}
\end{minipage}%
\caption{Distribution of $|\phi_h|$ (left plot) and correlation between $z$ and $|\phi_h|$ (right plot) for positive hadrons produced in exclusive events.}\label{fig:phih}
\end{figure}

The exclusive events are selected requiring only a pair of oppositely charged final state hadrons with total fractional energy $z_t=z_1+z_2>0.95$, and are clearly visibile in the $z_t$ distribution shown in the left plot of Fig. \ref{fig:zt}. The $z$ distribution for the positive hadrons of the pairs is shown in the right plot of the same figure. The broad structure at $0.4\lesssim z\lesssim 0.6$ is due to $\phi$ meson production, whose contribution is much smaller than that of the $\rho^0$.
Large $\cos\phi_h$ (and $\cos 2\phi_h$) modulations can be seen in the $|\phi_h|$ distribution of the exclusive hadrons, shown in the left plot of Fig. \ref{fig:phih} for $h^+$. Furthermore $|\phi_h|$ is strongly correlated with $z$, as can be seen from the right plot in Fig. \ref{fig:phih}, again for $h^+$. Indeed, the amplitude of the $\cos\phi_h$ modulation changes sign with $z$, due to momentum conservation in the decay of the exclusive VM. The same properties are observed also for $h^-$.

\begin{figure}[h]
	\centering
	\includegraphics[width=0.9\textwidth]{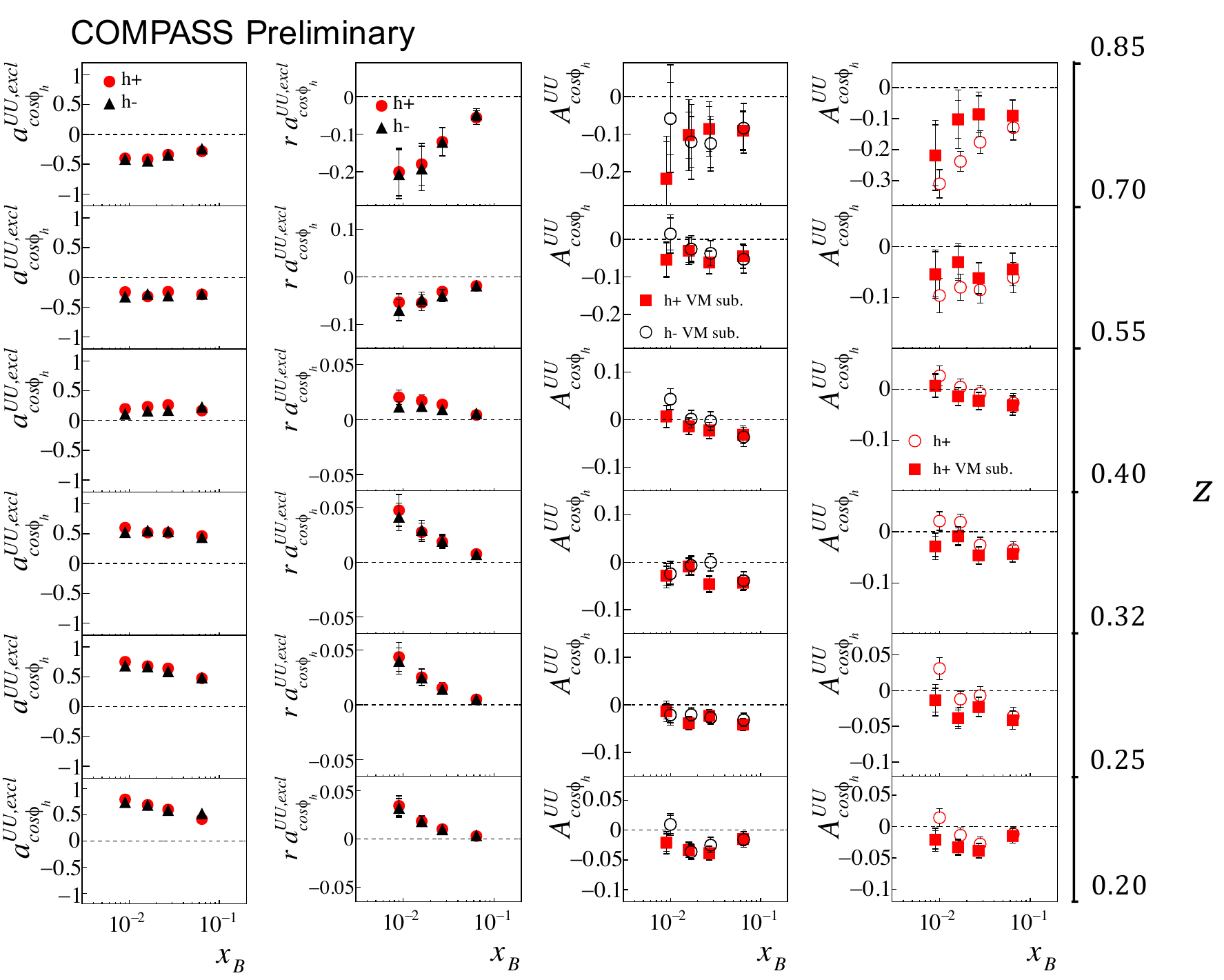}
\caption{First column: $a^{UU,excl}_{\cos\phi_h}$ amplitude for $h^+$ (circles) and $h^-$ (triangles). Second column: $r\,a^{UU,excl}_{\cos\phi_h}$ for $h^+$ (circles) and $h^-$ (triangles). Third column: $A^{UU}_{\cos\phi_h}$ asymmetry for $h^+$ (squares) and $h^-$ (circles) after the subtraction of exclusive VM contribution. Last column: comparison between the $A^{UU}_{\cos\phi_h}$ asymmetry before (open circles) and after exclusive VM subtraction (full squares) for $h^+$. All results are shown in the different $z$ bins and only for the smallest $p_T^h$ bin, namely $0.1\,\rm{GeV}/c<p_T^h<0.3\,\rm{GeV}/c$, of Ref. \cite{Adolph:2014pwc}.}
\label{fig:cosphi_all}
\end{figure}

\begin{figure}[h]
	\centering
	\includegraphics[width=0.9\textwidth]{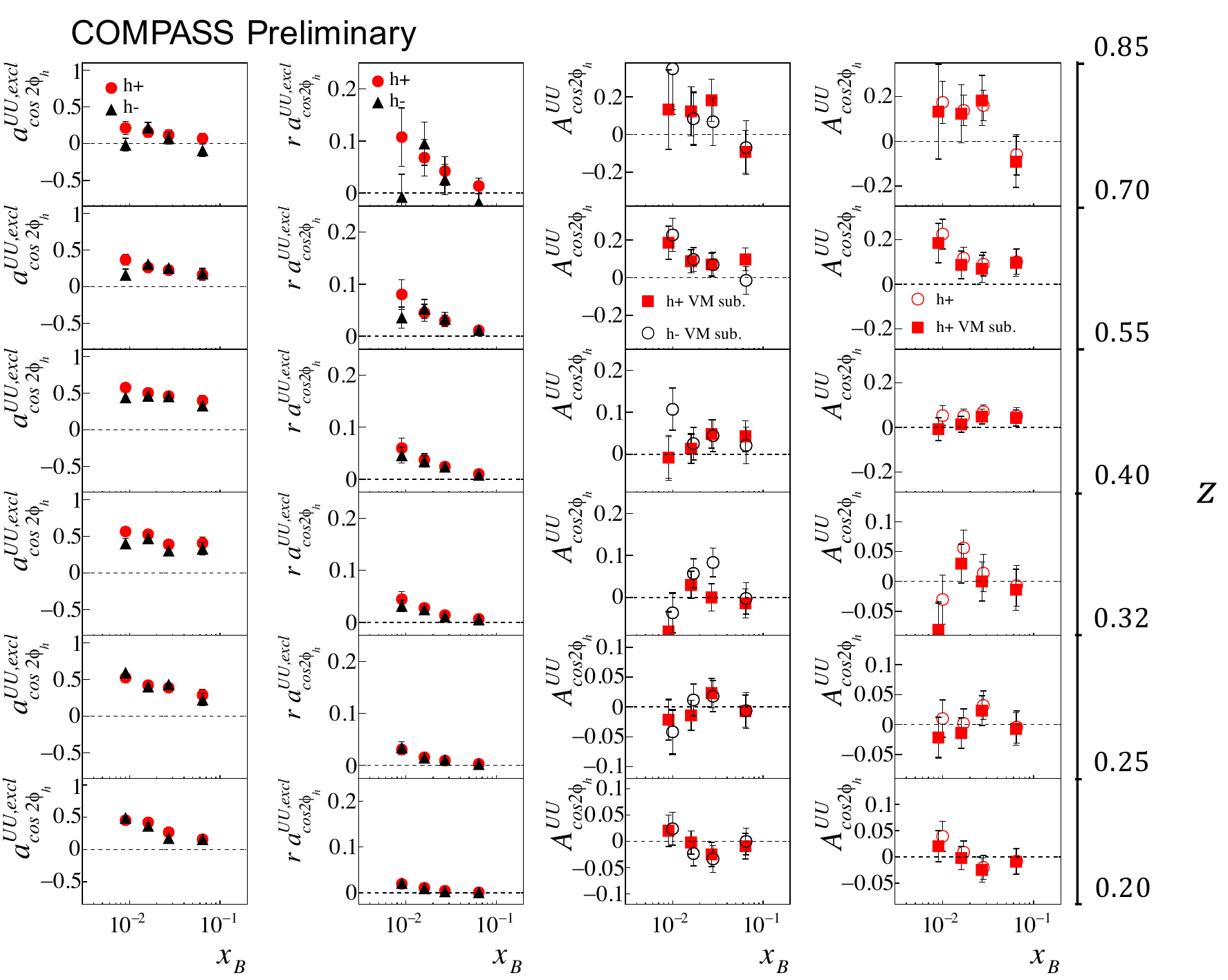}
\caption{First column: $a^{UU,excl}_{\cos 2\phi_h}$ amplitude for $h^+$ (circles) and $h^-$ (triangles). Second column: $r\,a^{UU,excl}_{\cos 2\phi_h}$ for $h^+$ (circles) and $h^-$ (triangles). Third column: $A^{UU}_{\cos 2\phi_h}$ asymmetry for $h^+$ (squares) and $h^-$ (circles) after the subtraction of exclusive VM contribution. Last column: comparison between the $A^{UU}_{\cos 2\phi_h}$ asymmetry before (open circles) and after exclusive VM subtraction (full squares) for $h^+$. All results are shown in the different $z$ bins and only for the smallest $p_T^h$ bin, namely $0.1\,\rm{GeV}/c<p_T^h<0.3\,\rm{GeV}/c$, of Ref. \cite{Adolph:2014pwc}.}
\label{fig:cos2phi_all}
\end{figure}

The amplitudes of the azimuthal modulations are extracted fitting the azimuthal distributions of exclusive hadrons using Eq. (\ref{eq:fexcl}). The resulting exclusive amplitudes $a^{UU,excl}_{\cos\phi_h}$ and $a^{UU,excl}_{\cos 2\phi_h}$ are shown in the first column of Fig. \ref{fig:cosphi_all} and in Fig. \ref{fig:cos2phi_all} respectively for both $h^+$ and $h^-$.
The $a^{UU,excl}_{\cos\phi_h}$ amplitude is very large at large and small $z$, and changes sign at $z\simeq 0.5$. The $a^{UU,excl}_{\cos 2\phi_h}$ amplitude is smaller but still non-negligible. Both amplitudes decrease with increasing $p_T^h$ and are almost equal for $h^+$ and $h^-$, indicating that what is modulated is the direction of the parent VM. For reasons of space, only the smallest $p_T^h$ bin of the analysis is shown in the figures.

\section{Percentage of hadrons from exclusive events}\label{sec:percentage}
For a quantitative estimate of the exclusive VM contribution to the unpolarized azimuthal asymmetries, the factor $r$ introduced in Eq. (\ref{eq:r}) is needed. Here we use a parametrization obtained from previous works \cite{Adolph:2016bga,Adolph:2016bwc,Aghasyan:2017ctw}, which were based on a combined use of HEPGEN and LEPTO. This study is restricted to exclusive $\rho^0$ production which gives the main contribution to the exclusive hadrons in the SIDIS sample.

The parametrization we use for the fraction of exclusive hadrons in the total data sample is
\begin{equation}\label{eq:r_parametrization}
    r(x_B,z,P_{\rm{T}})=A(x_B)\, e^{z\,B(x_B)}\,C(p_T^h,z)
\end{equation}
where the functions $A$ and $B$ are given by
\begin{eqnarray}\label{eq:A}
    A(Q^2)=\frac{0.004}{(Q^2-0.86)^{0.89}},
    && B(Q^2)=6.83\left(\frac{Q^2}{Q^2-0.74}\right)^{-0.43}
\end{eqnarray}
with $Q^2$ in $(\rm{GeV^2}/c^2)$, and the $x_B$ dependence of $Q^2$ parametrized as $\langle Q^2\rangle = 0.76+65.1\,\langle x_B\rangle +426.1\,\langle x_B\rangle^2$,
which is valid at COMPASS kinematics.

Also, the $p_T^{h}$ dependence of the quantity $r(x_B,z,p_T^{h})$ is encoded in a discrete function $C(p_T^{h},z)$, evaluated in each $p_T^{h}$ bin at fixed $z$ and normalized to unity
\begin{equation}
    \sum_i C(p_{Ti}^h,z)\Delta p_{Ti}^h=1
\end{equation}
where the sum runs over the different $p_T^h$ bins. The values used in this study are given in Tab. \ref{tab:NPT}.

\begin{table}
    \begin{tabular}{l|l|l|l|l}
    $p_T^h\, (\rm{GeV}/c)$ & $0.6<z<0.8$ & $0.4<z<0.6$ & $0.3<z<0.4$ & $0.2<z<0.3$ \\ \hline
    $0.10<p_T^h<0.30$& $1.67$ & $1.60$ & $1.98$ & $2.17$ \\ \hline
    $0.30<p_T^h<0.50$ & $1.62$ & $1.70$ & $1.39$ & $1.74$ \\ \hline
    $0.50<p_T^h<0.64$ & $1.18$ & $1.06$ & $0.79$ & $0.43$ \\ \hline
    $0.64<p_T^h<1.0$ & $0.49$ & $0.53$ & $0.60$ & $0.43$ \\
    \end{tabular}
    \caption{Values of $C(p_T^h,z)$ in the four $p_T^h$ bins
    of Ref. \cite{Adolph:2014pwc} for different $z$ intervals.}\label{tab:NPT}
\end{table}

\begin{figure}[h]
	\centering
	\includegraphics[width=0.9\textwidth]{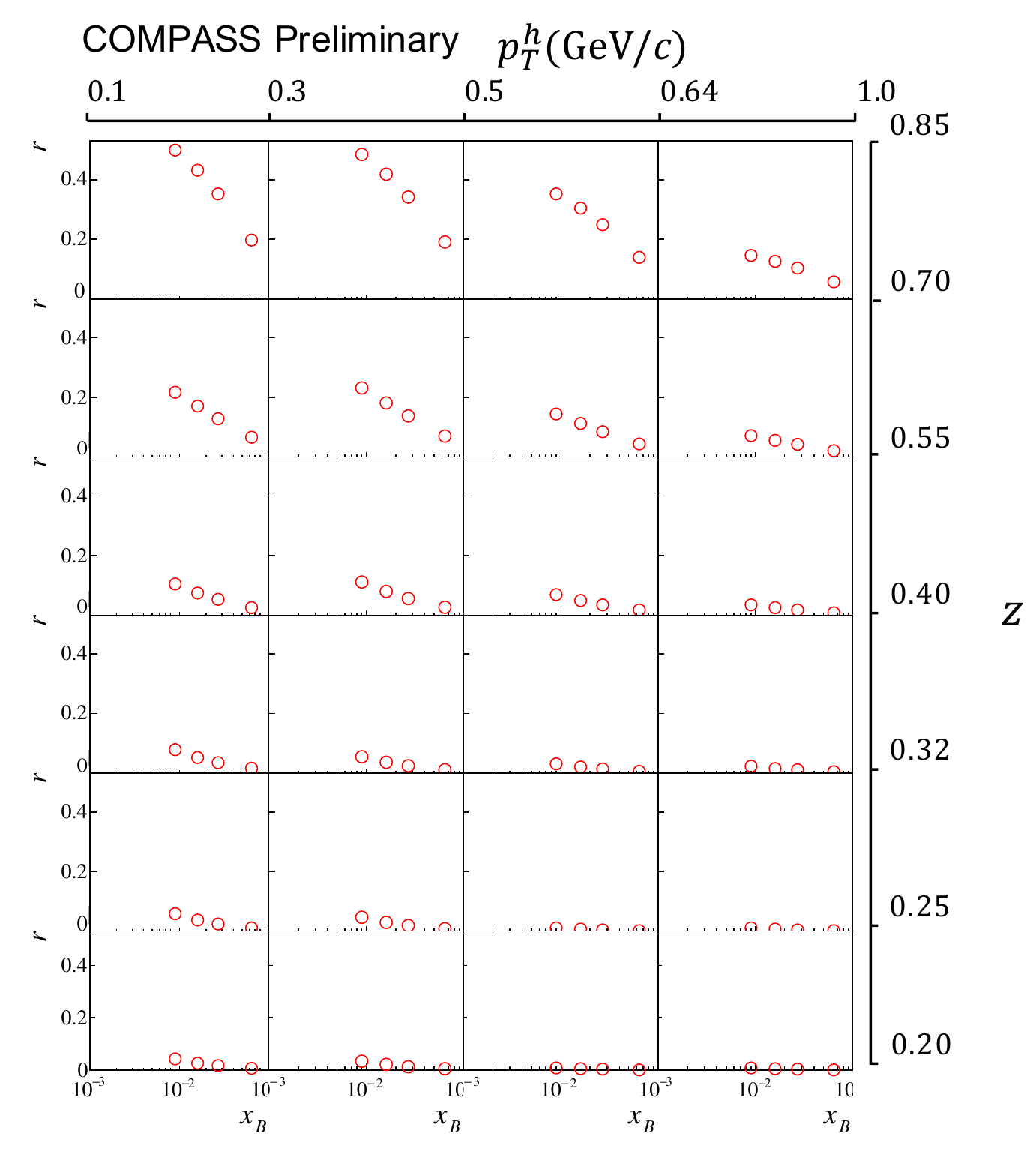}
\caption{Contamination coefficient $r$ evaluated as function of $x_B$ in the 3D binning of Ref. \cite{Adolph:2014pwc}.}
\label{fig:r}
\end{figure}

The final result for $r$ in all the 3D kinematical bins is shown in Fig. \ref{fig:r}.
As one can see the fraction of pions coming from the decay of exclusively produced $\rho^0$'s is very large at large $z$ and at small $p_T^h$, where it reaches $50\%$, and diminishes for decreasing $z$ and increasing $p_T^h$.
Note that in Refs. \cite{Adolph:2016bga,Adolph:2016bwc,Aghasyan:2017ctw} the quoted systematic uncertainty on $r$ is $30\%$ due to the uncertainty on the theoretical diffractive cross section.

\section{Unpolarized SIDIS azimuthal asymmetries after subtracting the contribution of exclusive VMs.}\label{sec:pureSIDIS}

The exclusive hadrons contributions to the published azimuthal asymmetries $r\,a^{UU,excl}_{\cos i\phi_h}$ of Eq. (\ref{eq:pure_A}) are obtained combining the contamination factor given in Sec. \ref{sec:percentage} with the exclusive amplitudes of Sec. \ref{sec:measurement}. The results are shown for both $h^+$ (circles) and $h^-$ (triangles) in the second column of Fig. \ref{fig:cosphi_all} and Fig. \ref{fig:cos2phi_all} for the $\cos\phi_h$ and $\cos 2\phi_h$ modulations respectively. Also in this case, only the results for the smallest $p_T^h$ bin, namely $0.1\,(\rm{GeV}/c) < \it{p_T^h}<0.3\,(\rm{GeV}/c)$, are shown. As can be seen, the contribution of exclusive hadrons is clearly different from zero in both modulations and reaches values up to $20\%$ at large $z$ and small $p_T^h$ for the $\cos\phi_h$ case. The contribution to the $\cos 2\phi_h$ modulation is smaller but still non-negligible, in particular if compared to the measured values of the asymmetries.

The measured azimuthal asymmetries corrected for the contamination of the exclusive vector mesons as in Eq. (\ref{eq:pure_A}) are shown in the third column of Fig. \ref{fig:cosphi_all} and of Fig. \ref{fig:cos2phi_all} respectively, again for the smallest $p_T^h$ bin. In particular the $A^{UU}_{\cos\phi_h}$ asymmetry is smoother after the subtraction and only a few positive points remain, which are hard to be described by the Cahn effect.

The last column of Fig. \ref{fig:cosphi_all} and of Fig. \ref{fig:cos2phi_all} shows the comparison between the asymmetries as published (open points) and after subtracting the contribution of exclusive VMs (full points) for $h^+$. In particular from Fig. \ref{fig:cosphi_all} we can see that the contribution of exclusive hadrons is sizeable also at small $z$. The effect is smaller for the $A^{UU}_{\cos 2\phi_h}$ asymmetry but still visible.

\begin{figure}[tbh]
	\centering
	\includegraphics[width=0.9\textwidth]{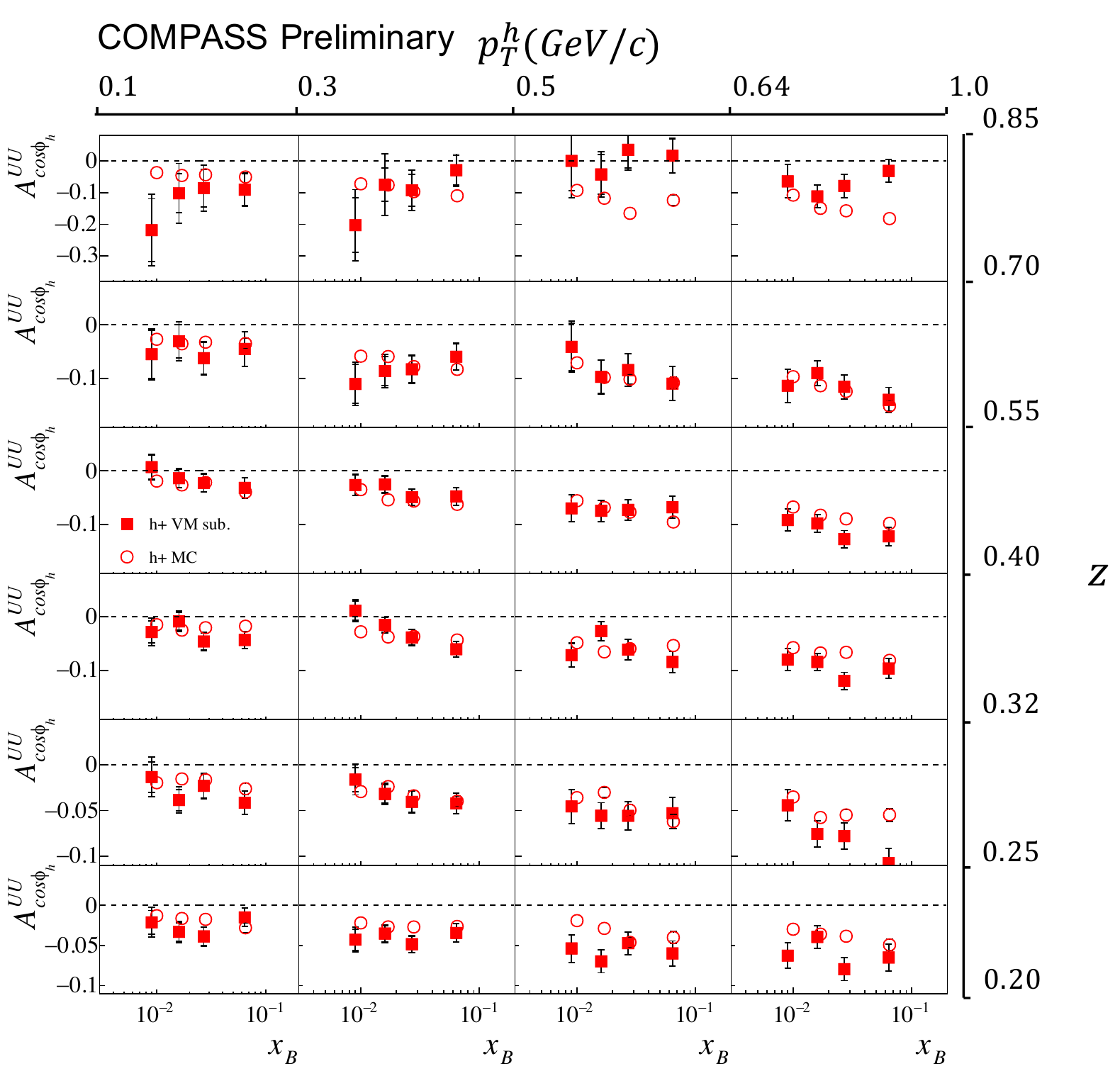}
\caption{Comparison between the SIDIS $A^{UU}_{\cos \phi_h}$ asymmetry after the subtraction of the exclusive VM contribution (squares) and MC (circles) for $h^+$.}
\label{fig:sub_MC_cosphi_pos}
\end{figure}

Finally in Fig. \ref{fig:sub_MC_cosphi_pos} the MC (open points) is compared with the SIDIS $A^{UU}_{\cos\phi_h}$ asymmetry for $h^+$ (full points) after the subtraction of the exclusive VM contribution in all the $p_T^h$ bins of our analysis. The agreement is good and the trends are similar over all bins, except for large $z$ and $p_T^h>0.5\,(\rm{GeV}/c)$.

\section{Conclusions}\label{sec:conclusions}
The COMPASS Collaboration has measured the azimuthal modulations of
positive and negative
hadrons from the decay of exclusive vector mesons produced in 
the scattering
of $160\,\,(\rm{GeV}/c)$ muons on a ${}^6LiD$ target. The amplitudes of the modulations are found to be large
and of the same sign for positive and negative hadrons.

These "exclusive" hadrons constitute a contamination to the SIDIS sample. Their contribution to the published COMPASS $A^{UU}_{\cos \phi_h}$ and $A^{UU}_{\cos 2\phi_h}$ unpolarized azimuthal asymmetries is estimated quantitatively and shown to be non-negligible over all the explored kinematical region and in particular at large $z$. After subtracting the $\cos \phi_h$ hadron amplitudes of the "exclusive" hadrons, the published COMPASS $A^{UU}_{\cos\phi_h}$ asymmetries turn out to be in reasonable 
agreement over most of the explored kinematic region with a Monte Carlo simulation implementing the Cahn effect, except for a very few bins at large $z$ and large $p_T^h$.

The results presented here are an important finding since the estimation of such contribution, in so far neither evaluated by COMPASS nor by HERMES, impact the phenomenological analyses which now could hopefully be successful and succeed in disentangling the various effects present in the data providing a first evidence for the Boer-Mulders effect.

\end{document}